\documentclass[12pt]{article}
\usepackage{epsfig}
\usepackage{url}
\usepackage[section]{placeins}
\usepackage{float}
\def\Title#1{\begin{center} {\Large {\bf #1} } \end{center}}

\begin{document}

\Title{Studies to Understand and Optimize the Performance of Scintillation Counters for
the Mu2e Cosmic Ray Veto System}

\bigskip\bigskip

%+\addtocontents{toc}{{\it D. Reggiano}}
%+\label{ReggianoStart}

\begin{raggedright}  

{\it Peter Farris, Craig Group, Yuri Oksuzian, Pedrom Zadeh\\
Department of Physics\\
University of Virginia\\
Charlottesville VA, USA}
\bigskip\bigskip
\end{raggedright}

\section{Introduction}

The Muon-to-Electron Conversion experiment (Mu2e) seeks to find physics beyond the Standard Model: the direct conversion of a muon to an electron without the emission of a neutrino ($\mu^- + N \rightarrow e^- + N$).  Such a transition violates the conservation of lepton numbers implied by the SM. To date, the best experimental limit on a direct muon-to-electron conversion is approximately $10^{-14}$ from the SINDRUM II experiment~\cite{Mesmer}. Mu2e expects to improve on this sensitivity by four orders of magnitude~\cite{Bartoszek:2014mya}.  

%to reach a single-event sensitivity of about $3.0 \times 10^{-17}$~\cite{Bartoszek:2014mya}.

\begin{figure}[htb]
\centering
\includegraphics[scale=0.12]{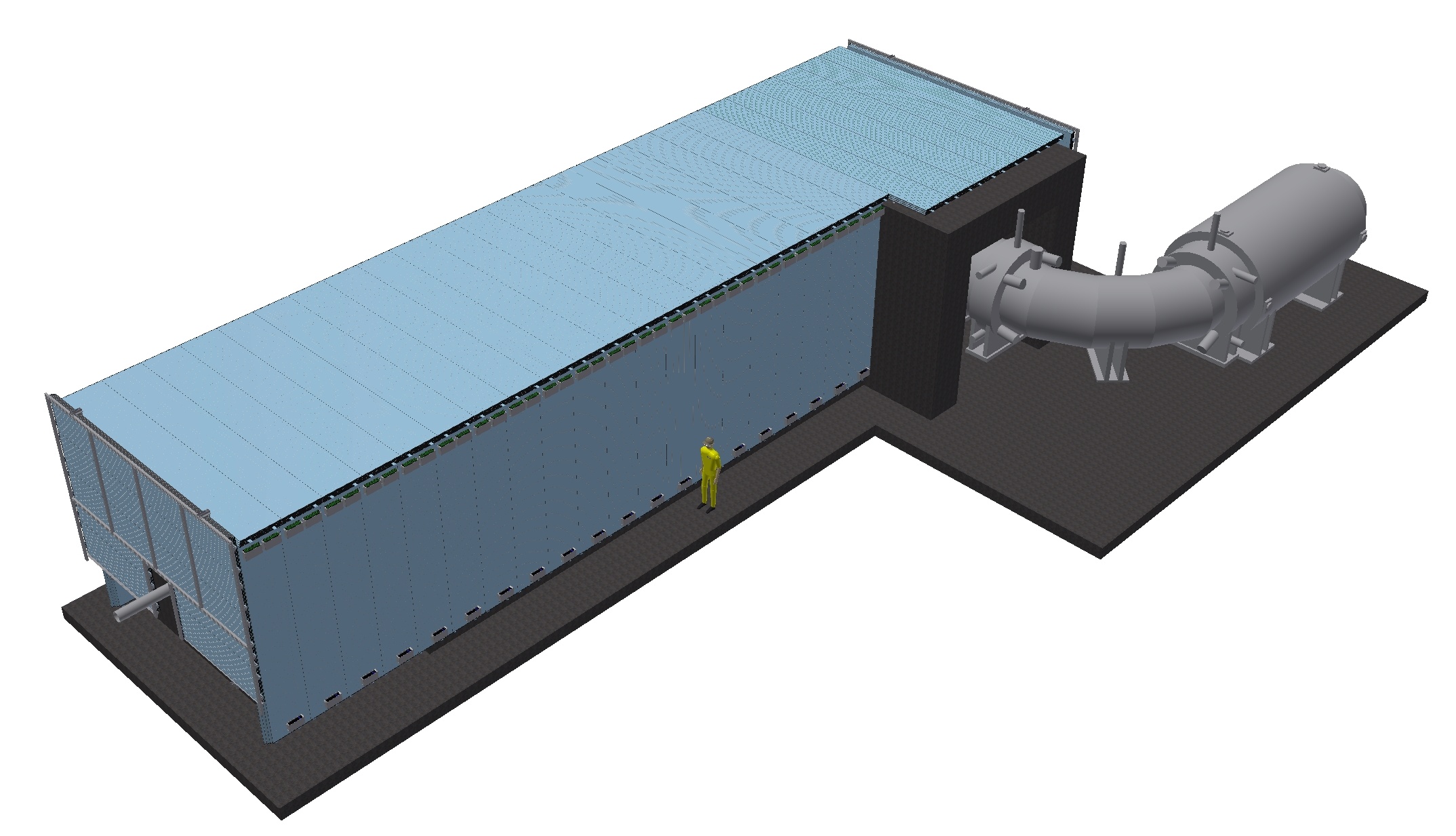}
\caption{A schematic of the CRV (in blue) enclosing the main detector (in grey).}
\label{fig:CRV_system}
\end{figure}

In order to reach the sensitivity goals, Mu2e must keep the total experimental background to well below one event.  Cosmic ray muons can interact with the material in the experiment and produce electrons that may mimic the desired signal.  The rate of these conversion-like backgrounds is about one per day, so Mu2e must reduce this background by several orders of magnitude to be successful.   A detector called the Cosmic Ray Veto (CRV) will surround the Mu2e experiment and veto these backgrounds with an efficiency of 99.99\% (see Fig.~\ref{fig:CRV_system}).  The CRV is comprised of dicounters, which are made by epoxying together two scintillation counters. Each dicounter is then outfitted with optical fibers (Kuraray double-clad Y11,non-S-type~\cite{ref:kuraray}) and instrumented with Hamamatsu silicon photomultipliers (SiPMs, model S13360-2050VE, 1584 pixels)~\cite{ref:hamamatsu}. A completed CRV module consists of four layers of dicounters (thirty-two dicounters total) epoxied to aluminum sheets which separate each layer. 

In order to optimize the performance of the CRV, \textit{Reflection Studies} and \textit{Aging Studies} were conducted. 

\section{Reflection Studies}
The objective of the reflection studies was to increase the light yield for the long scintillation counters with single-ended readout (counters with SiPMs only on one side).  An array of reflective materials were studied and their performance compared.

\subsection{Configuration}
The setup for the reflection studies consisted of a dicounter outfitted with SiPMs on one side and a given reflective material on the other. The dicounter was wrapped in light-tight aluminum foil and placed inside a light-tight box. A $1$ mCi Cesium-137 radioactive source was placed above the dicounter and used to produce scintillation within it. The SiPMs were connected to an electronics board that enabled the user to read out the current from the SiPM for each fiber channel. The following reflective materials were used: VM2000, silver Mylar, silver reflective tape, and SiPMs~\footnote{VM2000 is an enhanced specular reflector film sold by 3M~\cite{ref:vm2000}.}.

Each reflective material was placed at the far end of the dicounter, and light yield measurements were taken with and without the Cs source at 10~cm intervals from 20~cm to 430~cm from the readout end. In order to subtract the thermal noise in the SiPM, recorded current values \textit{without} the radioactive source (called dark current) were measured and subtracted.

\subsection{Theoretical Model and Analysis}
Assuming an equal probability for the light to go in either direction, the amount of light making it to the SiPMs directly from an arbitrary position $x$ can be approximated by
\begin{equation}
L=A(e^{-\frac{x}{\lambda(x)}}+fe^{-\frac{x'}{\lambda(x')}}),
\end{equation}
where $A$ is the initial amount of light produced, $\lambda(x)$ is the effective attenuation length as a function of position, $f$ is the reflection factor of a given material, and $x'$ is the distance the reflected light travels. Note that $x'=2\ell-x$ for $\ell=4.5\mathrm{m}$ being the length of the dicounter.

 The Cs source does not produce a single wavelength of light in the fibers, but rather a distribution of wavelengths ranging from about 450nm to 600nm.  The attenuation is wavelength dependent and has been measured for the CRV fiber~\cite{ref:DeZoort:2015pfn}.  As the light propagates in the fiber, the wavelength distribution changes, giving rise to a position dependency of the effective attenuation. Fig. 2 shows the light yield from one of the channels of the dicounter with its ends painted black. The data has been divided into four regions and exponential fits have been performed on each. For each fit, the average attenuation length over all four channels is measured. (All four channels exhibited the same trend).  Based on our measurements, the change in $\lambda$ is significant over the 4.5m length of the dicounter.

\begin{figure}[htb]
\centering
\includegraphics[scale=0.42]{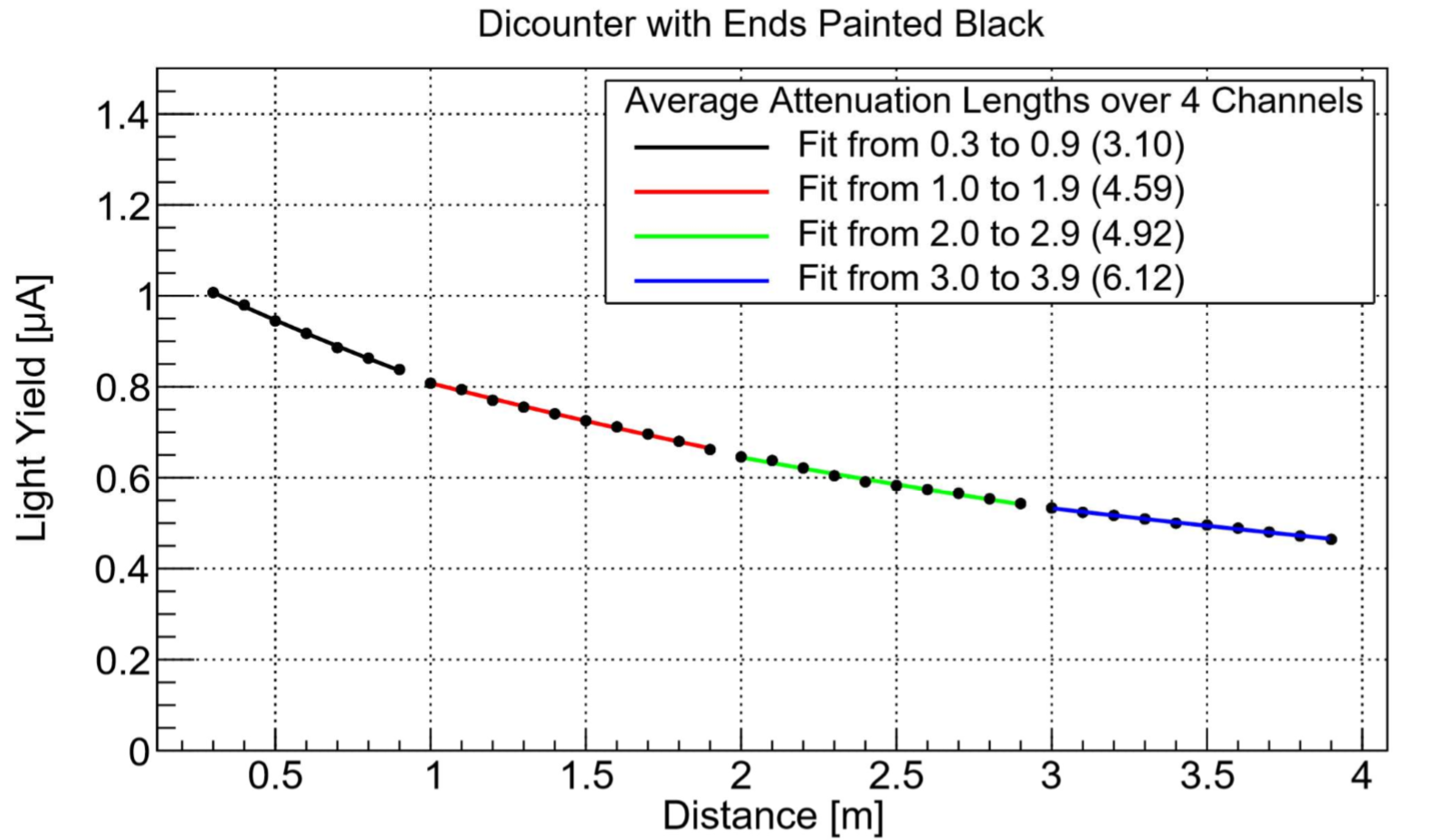}
\caption{Source light yield from the dicounter with ends painted black which shows the position-dependency of the effective attenuation length $\lambda$.}
\end{figure}

To better understand the improvements from a given reflector, each case was compared to the case with ends painted black (black paint was assigned $f=0$) by defining the ratio
\begin{equation}
R=\frac{L_{reflector}}{L_{black paint}}=1+fe^{\frac{x}{\lambda(x)}-\frac{x'}{\lambda(x')}}.
\label{eq2}
\end{equation}

Precise knowledge regarding $\lambda(x)$ was lacking. However, for small path length differences ($x \sim x'$), it seems reasonable to approximate $\lambda(x)$ with a linear function. Therefore, under the limit $x'\approx x$, the following convenient linear relationship was assumed
\begin{equation}
\lambda(x')=\frac{x'}{x}\lambda(x).
\label{assumption}
\end{equation}

This assumption is convenient because the argument of the exponential term in Eq.~\ref{eq2} is zero in this case.  For the above approximation to be reasonable, the measurement with the smallest path length
difference between the direct and reflected light was selected. Taking $x=4.3\mathrm{m}$, Eq. (2) was simplified to yield
\begin{equation}
f \sim R(4.3\mathrm{m})-1.
\end{equation}

From this simple relationship, the reflection factor for each of the materials was found.  An uncertainty of 7\% on the reflection factor extraction was estimated by considering limiting cases relative to the assumption in Eq.~\ref{assumption}. Table~\ref{tab:f} summarizes the measured reflection factors.

\begin{table}[H]
\begin{center}
\begin{tabular}{l|c}  
Reflective Material & $f \left(\pm 7\%\right)$ \\ \hline
VM2000 & 100 \\
Silver Mylar & 83 \\
Silver Tape & 84 \\
SiPMs & 20\\ \hline
\end{tabular}
\caption{Experimental reflection factors for each of the reflective material.}
\label{tab:f}
\end{center}
\end{table}
%%%%%%%%%%%%%%%%%%%%%%%%%%%%%%%%%%%%%%%%%%%%%%%%%%%%%%%%%%%%%%%%%%%%%%%%%%%

\begin{figure}[htb]
\centering
\includegraphics[scale=0.42]{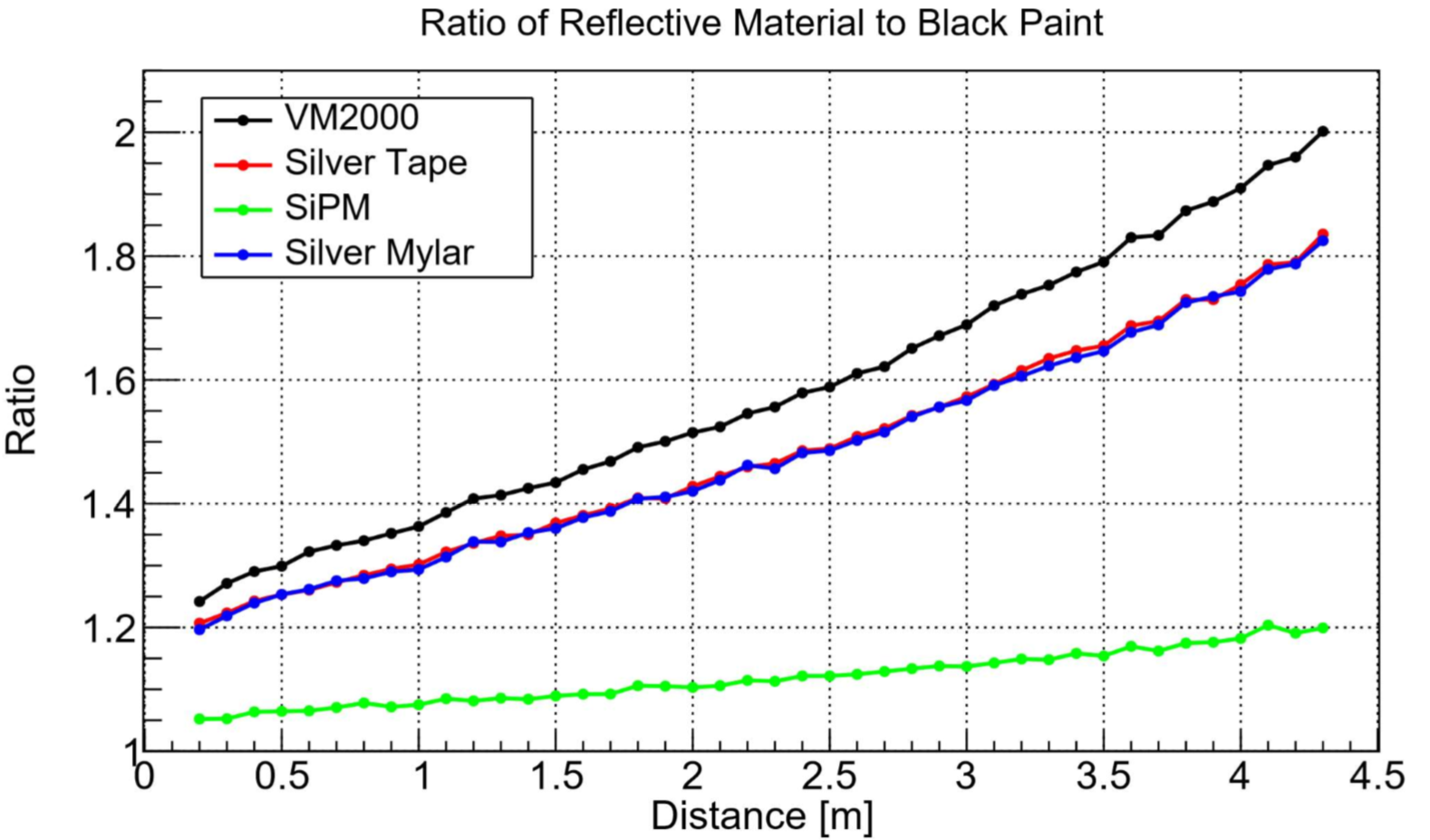}
\caption{Ratio of each reflective material to the case of black paint. The figure is from only one channel, but all channels exhibited the same trend. }
\label{fig:current_vs_distance}
\end{figure}

In Fig.~\ref{fig:current_vs_distance} the ratio of each of the reflectors to black paint for one channel is shown. This plot shows by how much light yield can be improved at each position along the counter depending on the material.  All channels exhibited the same trend.  The VM2000 had the highest reflection factor (consistent with 100\%).  Mylar, while not as good as VM2000, still reflects more than 80\% of the light.  Also notable was that the SiPM readout has a significant reflection factor of about 20\%. 

\section{Aging Studies}
In order to reduce the large background caused by cosmic ray muons, the CRV must maintain an efficiency of 99.99\% throughout the three-year lifetime of the experiment. However, natural degradation of the polystyrene scintillating counters could lower the efficiency of the CRV below the required value. The decay rate of the light yield of the scintillating counters was studied to determine if the required efficiency is maintained throughout the lifetime of the experiment.

\subsection{Theoretical Model}
The Arrhenius equation~\cite{ref:Arrhenius} was used to model the accelerated aging of the scintillating counters. The equation models the reaction rates as a function of temperature:
\begin{equation}
k(T)=Ae^{-\frac{E_a}{RT}},
\end{equation}
where $k(T)$ is the reaction rate constant, $A$ can be considered a constant, $E_a$ is the activation energy in kJ/mol, $T$ is the temperature, and $R$ is the universal gas constant. 

The ratio of two rate constants at different temperatures were used to find the effective aging time:
\begin{equation}
\frac{k_{high}}{k_{ref}}=e^{\frac{E_a}{R}\left(\frac{1}{T_{ref}}-\frac{1}{T_{high}}\right)},
\end{equation}
where $T_{ref}=23\mathrm{C}$, $T_{high}=50\mathrm{C}$, $E_a=91\mathrm{kJ/mol}$, $k_{ref}$ is the reference temperature rate constant~\cite{ref:EA}, and $k_{high}$ is the elevated temperature rate constant.  

\subsection{Configuration}
A total of twelve scintillating counters were used in this study--one control sample and eleven aging samples. The aging samples were placed in the oven and removed in four-week intervals. Thus, the twelve samples effectively modeled the aging of one sample through time.  After 11 samples had been removed, all samples were measured in a period of two days using the same removable fiber-SiPM jig (See Fig.~\ref{fig:aging}).

\begin{figure}[htb]
\centering
\includegraphics[scale=0.4]{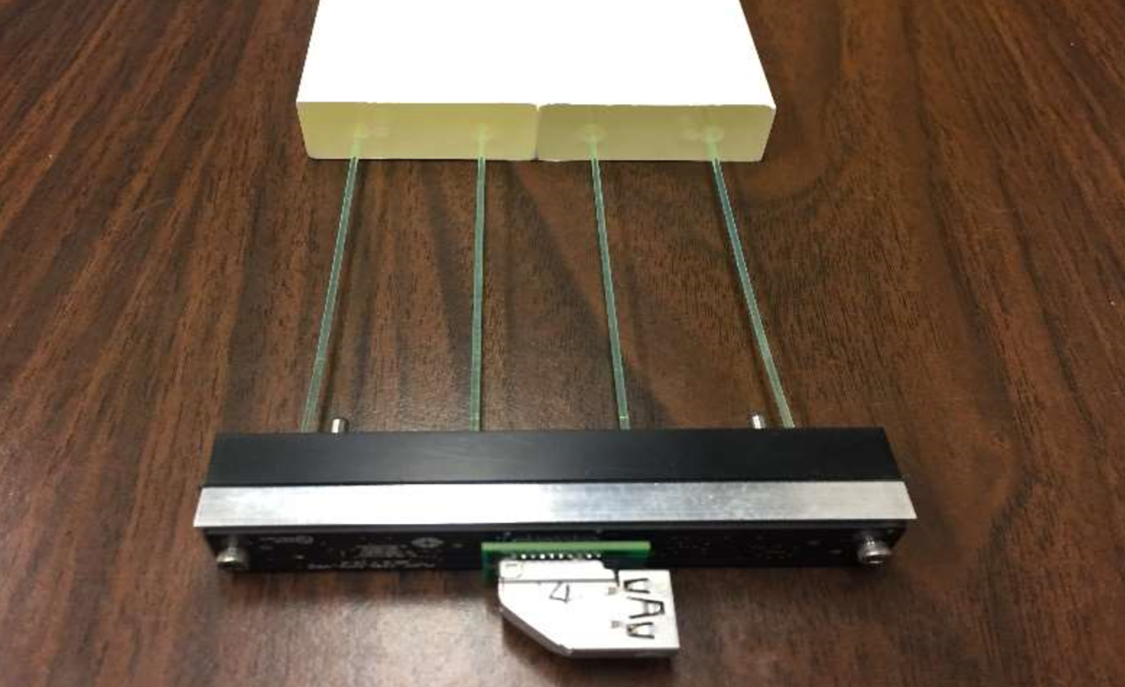}
\caption{Fiber-SiPM jig custom-made for the aging studies. This step ensured a degree of consistency between measurements.}
\label{fig:aging_setup}

\end{figure}

These methods ensured that aging of the measurement apparatus had no effect on the results. Additionally, the initial differences between each counter were within $\pm3\%$ and were corrected for in the final data. 

\subsection{Results}
\begin{figure}[htb]
\centering
\includegraphics[scale=0.4]{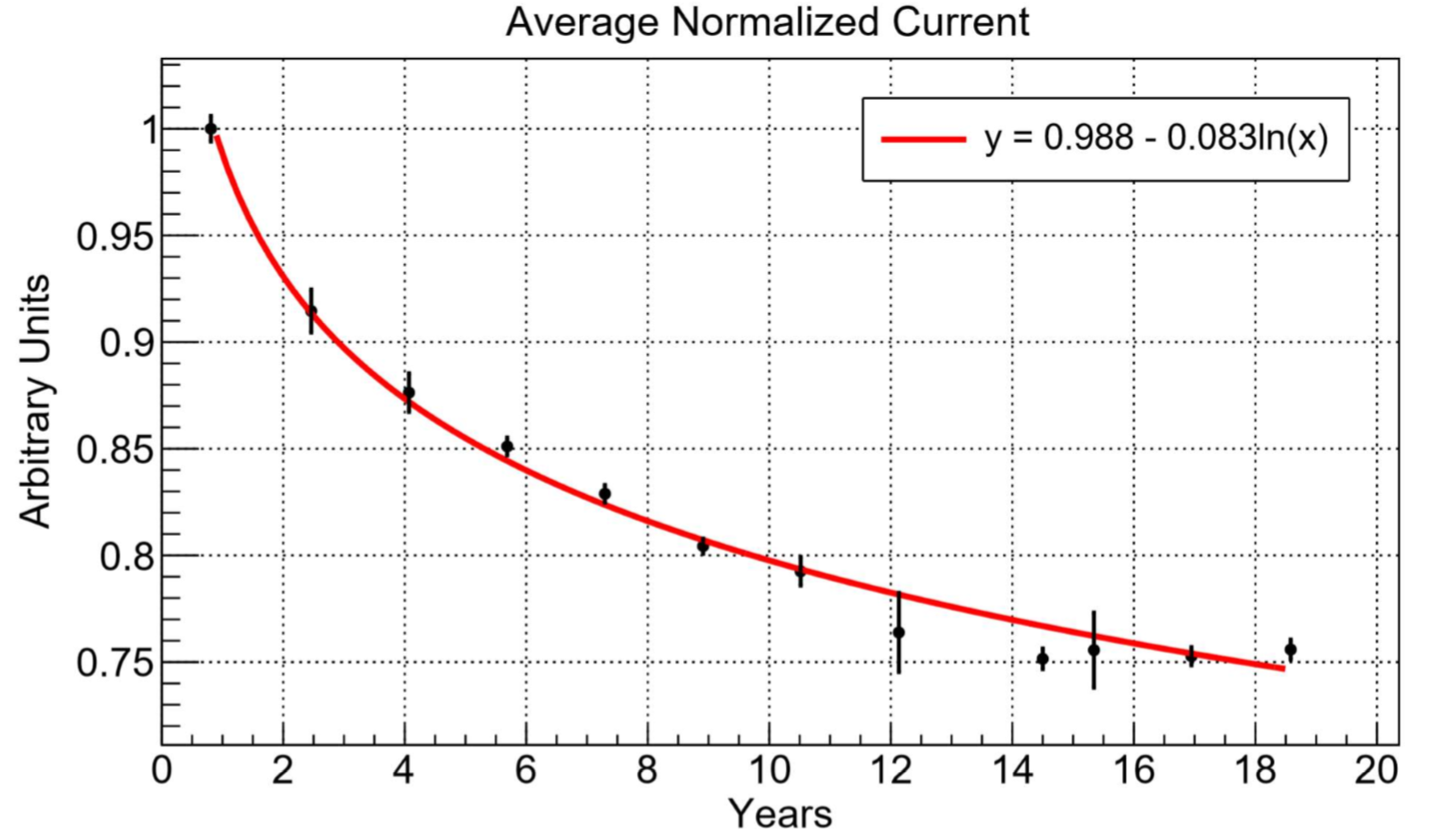}
\caption{Normalized net currents.}
\label{fig:aging_current}
\end{figure}

A sample measurement consisted of five runs. During each run, the thermal-background (dark) current and source current (produced from Cs-137) were measured five times and averaged for each of the four channels in the dicounter. The net current was found by subtracting the dark current from the source current and averaging over all five runs. The net currents for each channel were then normalized to their respective initial values and averaged. These values were then plotted against the effective aging time using the Arrhenius aging factor (See Fig.~\ref{fig:aging_current}).

\begin{figure}[htb]
\centering
\includegraphics[scale=0.4]{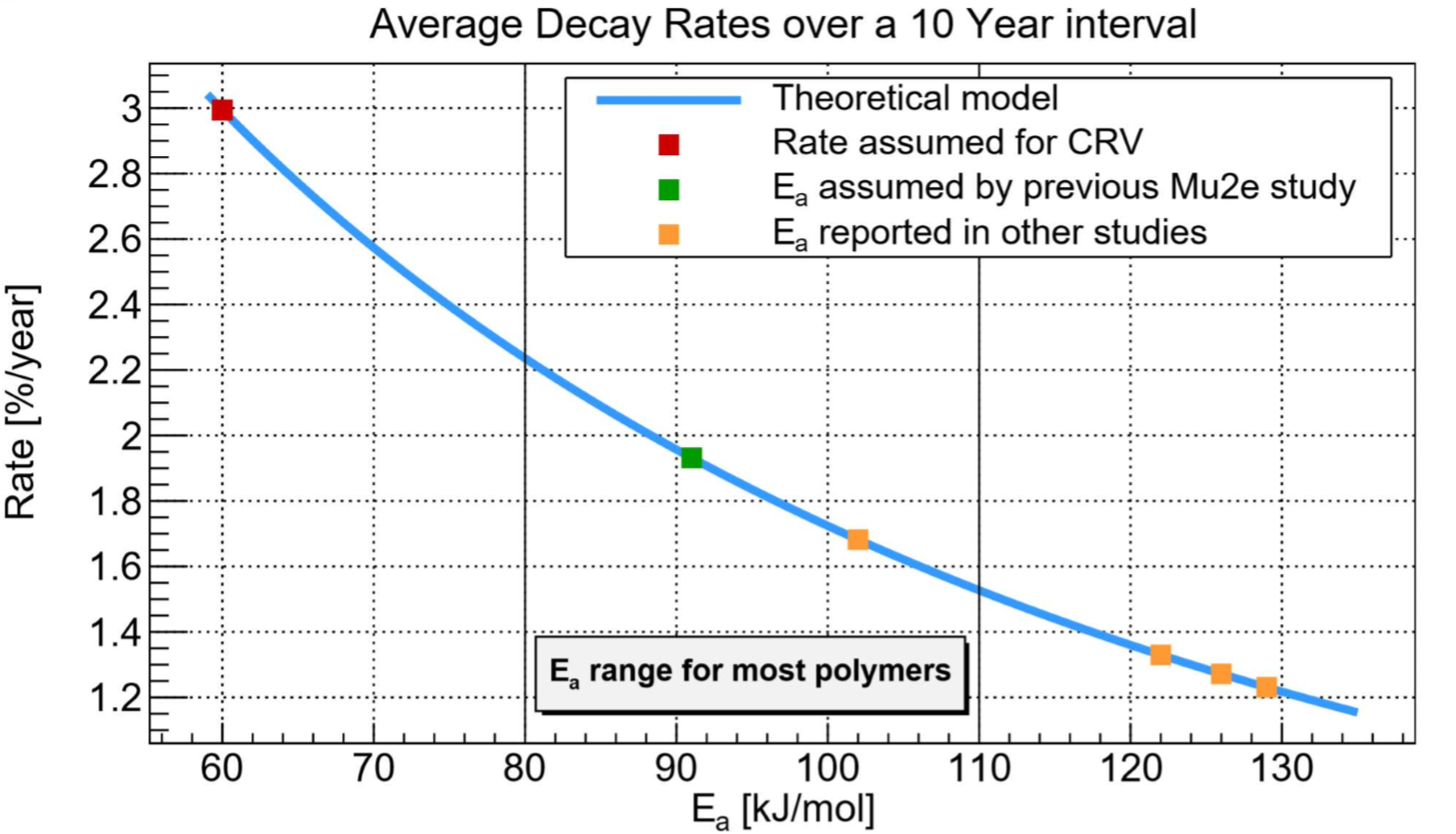}
\caption{Decay per year as a function of activation energy.}
\label{fig:aging}
\end{figure}

  One issue preventing an exact calculation of the aging rate is that the activation energy is not clearly defined. Other references provide values that differ from $E_a=91\mathrm{kJ/mol}$, the assumed value in a previous study~\cite{diCenzo, Muller}.  To compare the various values of activation energies,  Fig.~\ref{fig:aging} shows the decay per year over ten years as a function of activation energy. Also shown in Figure 6 is the assumed aging rate for the CRV at 3\% per year over ten years. As seen in Fig.~\ref{fig:aging}, this assumed rate is a conservative estimate.  Based on values for the activation energy found in the literature the aging rate for the CRV will be between 1.2\% and 1.9\% per year over ten years.

\section{Conclusions}
The Cosmic Ray Veto system plays a critical part in accounting for background muons in the Mu2e experiment. The ultimate goal of the two presented studies was to find ways to understand and optimize the response of the scintillating counters that are at the core of the CRV. While one study aimed to do so by increasing the light yield within each optical fiber, the other explored aging of the counters quantitatively to make sure the CRV would maintain the desired efficiency over time. Reflection studies showed promising results using VM2000 and the aging studies showed that the aging rate for the CRV scintillation counters should average $<3$\% per year over a ten-year time period .

\def\Discussion{
\setlength{\parskip}{0.3cm}\setlength{\parindent}{0.0cm}
     \bigskip\bigskip      {\Large {\bf Discussion}} \bigskip}
\def\speaker#1{{\bf #1:}\ }
\def\endDiscussion{}


\begin{thebibliography}{99}

%%
%%  bibliographic items can be constructed using the LaTeX format in SPIRES:
%%    see    http://www.slac.stanford.edu/spires/hep/latex.html
%%  SPIRES will also supply the CITATION line information; please include it.
%%

\bibitem{Mesmer}
W. Bertl et al, Eur. Phys. J. C47, 337 (2006).
%%CITATION = PWASA,13,1564;%%



%\cite{Bartoszek:2014mya}
\bibitem{Bartoszek:2014mya} 
  L.~Bartoszek {\it et al.} [Mu2e Collaboration],
  ``Mu2e Technical Design Report,''
  arXiv:1501.05241.
  %%CITATION = ARXIV:1501.05241;%%
  %83 citations counted in INSPIRE as of 18 Sep 2017


\bibitem{ref:kuraray}
Kuraray America, Inc., 200 Park Ave., NY 10166 USA; 3-1-6, NIHONBASHI, 
CHUO-KU, TOKYO 103-8254, JAPAN. 

\bibitem{ref:hamamatsu}
Hamamatsu Photonics K.K., 325-6 Sunayama-cho, Naka-ku, Hamamatsu City, 
Shizuoka Pref., 430-8587, Japan. Hamamatsu Corp., 360 Foothill Rd., 
Bridgewater, NJ 08807.


\bibitem{ref:vm2000}
Vikuiti™ Enhanced Specular Reflector Film,
3M, 3M Center, St.\ Paul, MN 55144, USA.


\bibitem{ref:DeZoort:2015pfn} 
  G.~DeZoort, E.~C.~Dukes, R.~C.~Group, H.~Kessenich, Y.~Oksuzian, T.~Rase and D.~Shooltz, 
  ``Performance of Wavelength-Shifting Fibers for the Mu2e Cosmic Ray Veto Detector,'' 
  arXiv:1511.06225. 
  %%CITATION = ARXIV:1511.06225;%%


\bibitem{ref:Arrhenius}
Arrhenius Equation, \textit{Encyclopedia Britannica Online}, \url{https://www.britannica.com/science/Arrhenius-equation}, April 24 2017.


\bibitem{ref:EA}
R.S.  Spencer and D.E. Dillon,  \textit{The viscous flow of molten polystyrene}, J. Colloid Sci., 3, 163-180, 1948.


\bibitem{diCenzo}
Jeffery D. Peterson, Sergey Vyazokvkin, Charles A. Wight, \textit{Kinetics of the Thermal and Thermo-Oxidative Degradation of Polystyrene, Polyethylene and Poly(propylene)}, 2001. 
%%CITATION = TAADD,23,2647;%%

\bibitem{Muller}
Serge Bourbigot, Jeffery W. Gilman, Charles A. Wilkie, \textit{Kinetic Analysis of the Thermal Degradation of Polystyrene--montmorillonite nanocomposite}, June 2004.







\end{thebibliography}
\end{document}